\documentstyle[amssymb,aps,12pt]{revtex}

\begin{document}
\draft
\author{O. B. Zaslavskii}
\address{Department of Mechanics and Mathematics, Kharkov V.N. Karazin's National\\
University, Svoboda\\
Sq.4, Kharkov 61077, Ukraine\\
E-mail: aptm@kharkov.ua}
\title{Entropy of semiclassical 2D dilaton black holes away from the Hawking
temperature}
\maketitle

\begin{abstract}
Recently we showed that in semiclassical 2D dilaton gravity the regularity
of a black hole horizon may be compatible with divergencies of
Polyakov-Liouville stresses on it, the temperature deviating from its
Hawking value. This makes the question about thermal properties of such
solutions non-trivial. We demonstrate that, adding to gravitation-dilaton
part of the action certain counterterms, which diverge on the horizon but
are finite outside it, one may achieve finiteness of the effective
gravitation-dilaton couplings on the horizon. This gives for the entropy $S$
the Bekenstein - Hawking value in the nonextreme case and $S=0$ in the
extreme one similarly to what happens to ''standard '' black holes.
\end{abstract}

\pacs{PACS numbers: 04.60Kz, 04.70.Dy}


\section{Introduction}

Two-dimensional (2D)\ dilaton theories of gravity \cite{callan} (for a
recent review see \cite{od}, \cite{dv}) attract much attention for a number
of reasons. In particular, they contain solutions of a black hole type that
gives a possibility to trace in details, exploiting simple models, rather
subtle effects of interaction between curvature and quantum fields, that in
the more realistic 4D case is obscured by mathematical complexities. In the
present Letter we touch upon only one aspect of this subject, concerning
connection between quantum backreaction and black hole thermodynamics.

It is known that, inasmuch as backreaction is neglected, the Euclidean
action and black hole entropy are finite at arbitrary temperature for both
the non-extreme and extreme cases. Once backreaction is taken into account,
this brings about severe restrictions into black hole thermodynamics. For
non-extreme black holes this enforces the choice of the temperature $T=T_{H}$%
, where $T_{H}=\kappa /2\pi $ is the Hawking temperature, $\kappa $ is the
surface gravity, i.e. a pure geometrical characteristics of a system.
Otherwise, for $T\neq T_{H}$, quantum stresses blow up on the horizon that,
typically, destroy a regular horizon completely. These circumstances were
traced in detailed in \cite{and}, where the exact form of 2D stress-energy
tensor was used explicitly. While the equality $T=T_{H}$ represents for
non-extreme black holes well-established connection between geometry,
quantum theory and thermodynamics, for the extreme case $T_{H}=0$ the
situation is not so obvious. It was suggested in \cite{ross} (see also \cite
{teit} - \cite{kum}) to ascribe an arbitrary temperature $T$ to extreme
black holes. Then one can calculate the Euclidean action and find the
entropy $S=0$. Such a prescription works well on a pure classical level but
quantum corrections, caused by backreaction, however small they be, destroy
this picture completely for the same reasons as in the non-extreme case.
Therefore, one is forced to put $T=T_{H}=0$, that makes thermodynamics
questionable.

In previous works we showed that there exist exceptional situations, when
infinite quantum backreaction on the horizon of dilaton black holes (in
contrast to general relativity) may be compatible with regular geometry of a
horizon both in the non-extreme \cite{nonext} and extreme \cite{ext} cases.
In doing so, infinite backreaction was in a sense compensated by infinite
gravitation-dilaton coupling and infinite dilaton gradients. Meanwhile, for
non-extreme black holes the entropy is proportional to the horizon value of
this coupling, so it formally diverges as well as the Euclidean action as a
whole. Thus, thermodynamic interpretation for such solutions fails. For
extreme black holes, considered in \cite{ext}, the entropy formally is zero,
but the potential problem, caused by an infinite coupling, is connected with
the energy, associated with the horizon since this energy is proportional to
its gradient.

The aim of this Letter is to pay attention that there exists a possibility
to combine the inequality $T\neq T_{H}$ (and, thus, infinite backreaction)
not only with the regularity of a horizon, but also with a well-defined
entropy. This circumstance is especially important for the extreme case
since it supplies us with examples, when the black hole entropy, whose very
notion is the subject of intensive discussion in recent years \cite{vaf} - 
\cite{claus}, may be advocated on a {\it semiclassical }(not only pure
classical) level.

\section{Non-extreme black holes and exactly solvable models}

Let us consider the system governed by the action 
\begin{equation}
I=I_{0}+I_{PL}\text{,}  \label{1}
\end{equation}
where 
\begin{equation}
I_{0}=\frac{1}{2\pi }\int_{M}d^{2}x\sqrt{-g}[F(\phi )R+V(\phi )(\nabla \phi
)^{2}+U(\phi )]+\frac{1}{\pi }\int_{\partial M}dskF(\phi )\text{.}  \label{2}
\end{equation}
Here the boundary term with the second fundamental form $k$ makes the
variational problem self-consistent, $ds$ is the line element along the
boundary $\partial M$ of the manifold $M.$

$I_{PL\text{ }}$is the Polyakov-Liouville action \cite{polyakov}
incorporating effects of Hawking radiation and its backreaction on the black
hole metric for a multiplet of N scalar fields. It is convenient to write it
down in the form \cite{solod}, \cite{fis} 
\begin{equation}
I_{PL}=-\frac{\kappa }{2\pi }\int_{M}d^{2}x\sqrt{-g}[\frac{(\nabla \psi )^{2}%
}{2}+\psi R]-\frac{\kappa }{\pi }\int_{\partial M}\psi kds\text{.}  \label{3}
\end{equation}
The function $\psi $ obeys the equation 
\begin{equation}
\Box \psi =R\text{,}  \label{4}
\end{equation}
where $\Box =\nabla _{\mu }\nabla ^{\mu }$, $\kappa =N/24$ is the quantum
coupling parameter.

Varying the action with respect to metric gives us $(T_{\mu \nu }=2\frac{%
\delta I}{\delta g^{\mu \nu }})$: 
\begin{equation}
T_{\mu \nu }^{(tot)}\equiv T_{\mu \nu }^{(0)}+T_{\mu \nu }^{(PL)}=0\text{,}
\label{6}
\end{equation}
where 
\begin{equation}
T_{\mu \nu }^{(0)}=\frac{1}{2\pi }\{2(g_{\mu \nu }\Box F-\nabla _{\mu
}\nabla _{\nu }F)-Ug_{\mu \nu }+2V\nabla _{\mu }\phi \nabla _{\nu }\phi
-g_{\mu \nu }V(\nabla \phi )^{2}\}\text{,}  \label{7}
\end{equation}
\begin{equation}
T_{\mu \nu }^{(PL)}=-\frac{\kappa }{2\pi }\{\partial _{\mu }\psi \partial
_{\nu }\psi -2\nabla _{\mu }\nabla _{\nu }\psi +g_{\mu \nu }[2R-\frac{1}{2}%
(\nabla \psi )^{2}]\}  \label{8}
\end{equation}
Variation of the action with respect to $\phi $ gives rise to the equation 
\begin{equation}
RF^{^{\prime }}+U^{\prime }=2V\Box \phi +V^{\prime }(\nabla \phi )^{2}\text{,%
}  \label{9}
\end{equation}
prime denotes derivative with respect to $\phi $.

Inasmuch as the auxiliary function $\psi $ can be expressed in terms of $%
\phi $ only, the action $I_{0\text{ }}$and the Polyakov-Liouville action are
combined in such a way that field equations (\ref{6}) - (\ref{8}) can be
formally obtained from the action $I_{0}$ only but with the ''renormalized''
coefficients which receive some shifts: $F\rightarrow \tilde{F},V\rightarrow 
\tilde{V}$ where 
\begin{equation}
\tilde{F}=F-\kappa \psi \text{,}  \label{10}
\end{equation}
\begin{equation}
\tilde{V}=V-\frac{\kappa }{2}\psi ^{\prime 2}\text{.}  \label{11}
\end{equation}
The dilaton equation (\ref{9}) can be rewritten, with account of (\ref{4}),
as 
\begin{equation}
R\tilde{F}^{^{\prime }}+U^{\prime }=2\tilde{V}\Box \phi +\tilde{V}^{\prime
}(\nabla \phi )^{2}\text{.}  \label{dil}
\end{equation}

Then there exists a class of exactly solvable models (see \cite{exact} and
literature quoted there), for which 
\begin{equation}
V=\omega (u-\frac{\kappa \omega }{2})  \label{v}
\end{equation}
or, equivalently, 
\begin{equation}
\tilde{V}=\omega \tilde{F}^{\prime }\text{, }\omega \equiv \frac{U^{\prime }%
}{U}\text{.}  \label{vt}
\end{equation}
Then it turns out that in the Schwarzschild gauge the metric of an eternal
black hole reads \cite{exact} 
\begin{equation}
ds^{2}=-fdt^{2}+f^{-1}dx^{2}\text{,}  \label{18}
\end{equation}
\begin{equation}
f=\frac{4\lambda ^{2}(\tilde{F}-\tilde{F}_{h})}{U}\text{, }x=\frac{\mu }{%
2\lambda }\text{, }\mu ^{\prime }=\tilde{F}^{\prime }e^{-\psi }\text{, }%
U=4\lambda ^{2}e^{\psi }\text{, }\psi ^{\prime }=\omega \text{,}  \label{g}
\end{equation}
the curvature 
\begin{equation}
R=\frac{U}{\tilde{F}^{\prime }}[\frac{U^{\prime }(\tilde{F}-\tilde{F}_{h)}}{U%
\tilde{F}^{\prime }}]^{\prime }\text{.}  \label{r}
\end{equation}

It was assumed in \cite{exact} that $\tilde{F}^{\prime }$ $\neq 0$ to ensure
the regularity of a horizon. Meanwhile, the equality $\tilde{F}=0$ may be
compatible with the regularity of a horizon if, at the same time, $U=0$.
Consider the case, when near the horizon 
\begin{equation}
\tilde{F}-\tilde{F}_{h}=A\left( \phi -\phi _{h}\right) ^{2}\text{,}
\label{fh}
\end{equation}

\begin{equation}
U=U_{1}(\phi -\phi _{h})\text{,}  \label{uh}
\end{equation}
where $A$ and $U_{1}$ are some constants.

It follows from (\ref{r}) that $R$ is finite. Near the horizon $\tilde{F}$
is finite, $\tilde{V}\backsim 2A$ is finite, $T_{\mu }^{\nu (tot)}$ are
finite. It is instructive to note that for our exactly solvable models 
\begin{equation}
\frac{\partial f}{\partial x}=2\lambda [1-\frac{(\tilde{F}-\tilde{F}%
_{h})\psi ^{\prime }}{\tilde{F}^{\prime }}]\text{,}
\end{equation}
\begin{equation}
T_{H}=\frac{1}{4\pi }\lim_{x\rightarrow x_{h}}\frac{\partial f}{\partial x}=%
\frac{\lambda }{\pi }\text{.}  \label{th2}
\end{equation}
As a result, our Hawing temperature is as twice as little as compared to the
standard case \cite{exact}, when $T_{H}=\frac{\lambda }{2\pi }$. We have
also 
\begin{equation}
\frac{\partial \tilde{F}}{\partial x}=(2\lambda )^{-1}U\text{, }\tilde{V}=%
\tilde{F}^{\prime }\psi ^{\prime }\text{.}
\end{equation}

Consider, for example, the spatial component. Then direct calculations give
us that $2\pi T_{1}^{1(tot)}=0$, as it should be. The gravitation-dilaton
part of the field equations 
\begin{equation}
2\pi T_{1}^{1(0)}=\frac{\partial f}{\partial x}\frac{\partial F}{\partial x}%
-U+Vf\left( \frac{\partial \phi }{\partial x}\right) ^{2}\text{,}
\end{equation}
near the horizon behaves like 
\begin{equation}
\frac{\pi T_{1}^{1(0)}}{2\lambda ^{2}}=\frac{3U_{1}}{8A}\kappa (\phi -\phi
_{h})^{-1}+...
\end{equation}

In a similar way, 
\begin{equation}
\frac{\pi T_{1}^{1(PL)}}{2\lambda ^{2}}=-\kappa (\frac{f}{2}\frac{\partial
\psi }{\partial x}+\frac{\partial f}{\partial x})\frac{\partial \psi }{%
\partial x}\text{.}  \label{pl1}
\end{equation}

Near the horizon 
\begin{equation}
\frac{\pi T_{1}^{1(PL)}}{2\lambda ^{2}}=-\kappa \frac{3U_{1}}{8A}(\phi -\phi
_{h})^{-1}+...  \label{pldiv}
\end{equation}
We see that divergent parts of gravitation-dilaton and Polyakov-Liouville
parts mutually compensate each other, as it should be. Thus, the entropy (up
to the constant) $S=2\tilde{F}_{h}$ \cite{action} is finite; $\tilde{F}$, $%
\tilde{V}$ are finite on the horizon, $S$ is well defined. The ''bare''
quantities $F$, $V$ diverge: according to (\ref{10}), (\ref{g}), (\ref{uh}),
near the horizon 
\begin{equation}
F=\tilde{F}+\kappa \ln (\phi -\phi _{h})+const\text{,}  \label{F}
\end{equation}
\begin{equation}
V=\tilde{V}+\frac{\kappa }{2}(\phi -\phi _{h})^{-1}\text{.}
\end{equation}
As $T_{\mu }^{\nu (PL)}$ diverges, it means in turn that $T\neq T_{H}$.
Indeed, this component may be represented as (see, e.g. \cite{and} for
details) $T_{1}^{1(PL)}=-\frac{\pi N}{6f}[T^{2}-\left( \frac{1}{4\pi }\frac{%
\partial f}{\partial x}\right) ^{2}]$. If spacetime is flat at infinity, $%
\frac{\partial f}{\partial x}\rightarrow 0$, and the parameter $T$ has the
meaning of temperature. Near the horizon, where $\frac{1}{4\pi }\frac{%
\partial f}{\partial x}\rightarrow T_{H}$, only the choice $T=T_{H}$ makes $%
T_{1}^{1(PL)}$ regular there. If $T_{1}^{1(PL)}$ diverges on the horizon,
this is inconsistent with the equality $T=T_{H}$.

\section{Extreme case}

Exactly solvable models, exploited for the analysis of the non-extreme case,
are now unsuitable since they give $T_{H}\neq 0$ according to (\ref{th2}).
Fortunately, the essence of matter becomes clear even without resorting to
exactly solvable models. Basic equations for $00$ and $11$ components read:

\begin{equation}
2f\frac{\partial ^{2}\tilde{F}}{\partial x^{2}}+\frac{\partial f}{\partial x}%
\frac{\partial \tilde{F}}{\partial x}-U-\tilde{V}f\left( \frac{\partial \phi 
}{\partial x}\right) ^{2}=0\text{,}  \label{sc00}
\end{equation}

\begin{equation}
\frac{\partial f}{\partial x}\frac{\partial \tilde{F}}{\partial x}-U+\tilde{V%
}f\left( \frac{\partial \phi }{\partial x}\right) ^{2}=0\text{.}
\label{sc11}
\end{equation}
It is also convenient to take the difference of (\ref{sc00}), (\ref{sc11})
to get 
\begin{equation}
\frac{\partial ^{2}\tilde{F}}{\partial x^{2}}=\tilde{V}\left( \frac{\partial
\phi }{\partial x}\right) ^{2}\text{.}  \label{dif}
\end{equation}

Let us consider extreme black holes for which $\tilde{F}$ and $\tilde{V}$
are finite on the horizon. Let also on the horizon $x=x_{h}$ the dilaton $%
\phi =\phi _{h}$ and $\frac{\partial x}{\partial \phi }\,$be finite. Then it
follows from (\ref{sc11}) that $U(\phi _{h})=0$. For a static metric eq. (%
\ref{4}) gives us 
\begin{equation}
\frac{\partial \psi }{\partial x}=\frac{a-\frac{\partial f}{\partial x}}{f}%
\text{.}
\end{equation}
Then it follows from (\ref{pl1}) that 
\begin{equation}
2\pi T_{1}^{1(PL)}=-\frac{\kappa }{2}\frac{(a+\frac{\partial f}{\partial x}%
)(a-\frac{\partial f}{\partial x})}{f}\text{.}  \label{plex}
\end{equation}
If $a=0$, the quantity $T_{1}^{1(PL)}$ is finite on the extreme horizon
since the denominator and numerator have the same order $(x-x_{h})^{2}$.
Let, however, $a\neq 0$. Then near the horizon$\frac{\partial \psi }{%
\partial x}\,=\frac{A_{1}}{(x-x_{h})^{2}}$ and 
\begin{equation}
\psi =-\frac{A_{1}}{x-x_{h}}=-\frac{A_{2}}{\phi -\phi _{h}}\text{,}
\end{equation}
$A_{2}=A_{1}\lim_{x\rightarrow x_{h}}\left( \frac{\partial \phi }{\partial x}%
\right) $. As, by assumption, $\tilde{F}$ is finite, the quantity $F=\tilde{F%
}+\kappa \psi $ behaves like $(x-x_{h})^{-1}\backsim (\phi -\phi _{h})^{-1}$%
, $V=\tilde{V}+\frac{\kappa }{2}\left( \frac{\partial \psi }{\partial \phi }%
\right) ^{2}$ behaves like $(\phi -\phi _{h})^{-4}$. (Note that quantities $%
F $ and $V$ should diverge in any case, if we want to have finite $\tilde{F}$
and $\tilde{V}$ but for $a=0$ divergencies are milder: $\psi \backsim \ln
(\phi -\phi _{h})$, $F\backsim \ln (\phi -\phi _{h})$, $V\backsim (\phi
-\phi _{h})^{2}$.) On one hand, eqs. (\ref{sc00}) and (\ref{dif}) have the
solutions which can be obtained by Taylor series in $(\phi -\phi _{h})\,$or $%
(x-x_{h})$, with finite $\tilde{F}_{h}$ and $\tilde{V}_{h}$. On the other
hand, for $a\neq 0$, $T_{1}^{1(PL)}$ diverges on the horizon according to (%
\ref{plex}). Direct calculations show that $T_{1}^{1(0)}$ have the leading
divergent term that compensates exactly that in (\ref{plex}), as it should
be. Thus, we get the regular solutions different parts of which diverge. The
total Euclidean action of the quantum-corrected system is well defined since 
$\tilde{F}$ is finite on the horizon (in this situation the enregy, as
usual, is due to the term on the physical boundary and does not contain the
contribution from the horizon). Then all the arguments of \cite{ross} - \cite
{kum} apply, with the only change that the classical coupling $F$ should be
replaced now by the quantum-corrected one $\tilde{F}$. Thus, the value $S=0$
is advocated in that case. In other words, we should introduce appropriate
counterterms into the gravitation-dilaton part of the action in such away,
that they diverge on the horizon. Then, if they behave like we described it
above, they compensate divergencies in Polyakov-Liouville part and give the
quite definite well-defined answer for the entropy. The divergencies in $%
I_{0}$ is an inevitable price for the finiteness of the total action. There
is no need to adjust very special logarithmic dependence for $\tilde{F}$ to
obtain semiclassical regular extreme black holes, as was done in \cite{ext}
- now $\tilde{F}$ is finite on the horizon.

\section{Summary}

A natural way to understand better fundamentals of black hole thermodynamics
is to try to violate some ''obvious'' assumptions, which are usually tacitly
assumed and remain unspoken, and to look at the consequences to which this
violation can lead$.$ As a result, we can either realize, why these
assumptions were necessary or, otherwise, extend some basic notions beyond
their original region of validity. In this Letter we abandoned the
assumption of the finiteness of the action coefficients on the horizon in
the bare gravitation-dilaton part. We demonstrated, how divergencies in the
corresponding couplings are connected with the violation of the condition $%
T=T_{H}$ and showed that this can be compatible with the well-defined black
hole entropy. In particular, for corresponding solutions the prescription
for the entropy value $S=0$ of extremal black holes \cite{ross} can be
advocated.

From the other hand, the procedure under discussion cannot be considered
only as a methodical exercise for elucidating, what happens if some
''obvious'' assumption about the bare Lagrangian are violated. The point is
that for a quantum-corrected system it is just quantities like $\tilde{F}$
and $\tilde{V}$ have direct meaning - meanwhile, they are finite in our
case. (To some extent, the relationship between $F$, $V$ and $\tilde{F}$, $%
\tilde{V}$ can be considered in our case as some analogue of the
renormalization procedure, but with the wording that this
''renormalization'' should remove divergencies only on the horizon - outside
the horizon all quantities are finite separately.) Only if one wishes to
trace back the behavior of different parts in field equation and splits them
explicitly to gravitation-dilaton part and quantum backreaction, the thermal
divergencies on the horizon come into play (otherwise they remain hidden,
since geometry near the horizon is perfectly smooth). This means that if
spacetime is asymptotically flat, at infinity $T_{\mu }^{^{\nu
}(PL)}\rightarrow \frac{\pi T^{2}N}{6}diag(1,-1)$ has the form, inherent to
thermal radiation, but with temperature $T\neq \kappa /2\pi $.

Combining the results of the previous works \cite{nonext}, \cite{ext} with
the present ones, we see that dilaton gravity gives us a more rich set of
possibilities than general relativity in what concerns thermal properties of
black holes and relationship between the character of quantum stresses on
the horizon and the regularity of the geometry.

For convenience, different cases are brought together in a table, where $"+"$
means ''exists'', while $"-"$ means ''does not exists'', the row ''0''
represents the classical case, all other rows refer to the semiclassical
situation. Here case 1 can be called typical and in fact discussion in \cite
{and} applies to it, case 1' showing, what happens if deviation of the
temperature from the Hawking value occurs. Case 2 refers to models of \cite
{nonext}, \cite{ext}, case 3 is the subject of the present article. Case 1
for extreme black hole means that $T=T_{H}$, so a regular horizon is
possible but, as $T_{H}=0$, there is no sensible thermodynamics. That for
extreme semiclassical black holes there two possibilities in case 2, is
explained by the fact that thermodynamics fails if $\tilde{F}$ and,
correspondingly, the energy associated with the horizon grow too rapidly
near the horizon (cf. \cite{inf}).

\newpage

Non-extreme black holes

\medskip

\begin{tabular}{|c|c|c|c|c|c|c|}
\hline
& $T$, $T_{H}$ & $F_{h},V_{h}$ & $\tilde{F}_{h},\tilde{V}_{h}$ & $T_{\mu
}^{\nu (PL)}$ on horizon & Regular horizon & Thermodynamics \\ \hline
0 & $=\,$or $\neq $ & finite & $\tilde{F}=F$, $\tilde{V}=V$ & $0$ & $+$ & $+$
($S=2F_{h}$) \\ \hline
1 & $=$ & finite & finite & finite & $+$ & $+$ ($S=2\tilde{F}_{h}$) \\ \hline
1' & $\neq $ & finite & infinite & infinite & $-$ & $-$ \\ \hline
2 & $\neq $ & infinite & infinite & infinite & $+$ & $-$ \\ \hline
3 & $\neq $ & infinite & finite & infinite & $+$ & $+$ ($S=2\tilde{F}_{h}$)
\\ \hline
\end{tabular}

\bigskip

\medskip

Extreme black holes

\medskip

\begin{tabular}{|l|l|l|l|l|l|l|}
\hline
& $T$ , $T_{H}=0$ & $F_{h},V_{h}$ & $\tilde{F}_{h},\tilde{V}_{h}$ & $T_{\mu
}^{\nu (PL)}$ on horizon & Regular horizon & Thermodynamics \\ \hline
0 & $\neq $ & finite & $\tilde{F}=F$, $\tilde{V}=V$ & $0$ & $+$ & $+$ ($S=0$)
\\ \hline
1 & $=$ & finite & finite & finite & $+$ & $-$ \\ \hline
1' & $\neq $ & finite & infinite & infinite & $-$ & $-$ \\ \hline
2 & $\neq $ & infinite & infinite & infinite & $+$ & $+$ ($S=0$) or $-$ \\ 
\hline
3 & $\neq $ & infinite & finite & infinite & $+$ & $+$ ($S=0$) \\ \hline
\end{tabular}

\bigskip

It is seen from the table that in 2D dilaton gravity black hole
thermodynamics becomes, on one hand, more restricted in that it fails to
exist at all in certain dilaton theories. However, from the other hand, it
extends to a more vast region than usually since it may withstand the
violation of the property $T=T_{H}$ (that manifests that an intimate link
between geometry and thermodynamics is now broken) and the appearance of
infinite stresses on the horizon. In the given context it is worth
mentioning that black hole thermodynamics fails to be mandatory also in some
4D theories with a scalar field (dilaton), even in the absence of quantum
effects, because of an infinite area of an event horizon or divergencies in
the coupling $F$; if $F$ grows too rapidly, the Euclidean action turns out
to be infinite \cite{inf}. Thus, divergencies in the gravitation-dilaton
coupling do not deprive the corresponding models of physical sense but,
rather, give rise to unusual and non-trivial relationship between the
existence of thermal properties, quantum backreaction and/or concrete nature
of the aforementioned divergencies.





%
%

%
%

\end{document}